\documentclass[aps,prb,reprint,twocolumn,superscriptaddress,showpacs]{revtex4-1}
\usepackage{graphicx}
\usepackage{latexsym}
\usepackage{amssymb}
\usepackage{amsmath}
\usepackage{amsfonts}
\usepackage{bm}
\usepackage{multirow}
\usepackage{color}

\begin{document}

\title{Natural orbitals renormalization group approach to a spin-$\frac{1}{2}$ impurity interacting with two helical liquids}

\author{Ru Zheng}
\email{zhengru@ruc.edu.cn}
\affiliation{Department of Physics, School of Physical Science and Technology, Ningbo University, Ningbo 315211, China}

\author{Rong-Qiang He}
\email{rqhe@ruc.edu.cn}
\affiliation{Department of Physics, Renmin University of China, Beijing 100872, China}

\author{Zhong-Yi Lu}
\email{zlu@ruc.edu.cn}
\affiliation{Department of Physics, Renmin University of China, Beijing 100872, China}

\date{\today}

\begin{abstract}
Using the natural orbitals renormalization group, we studied the problem of a localized spin-$\frac{1}{2}$ impurity coupled to two helical liquids via the Kondo interaction in a quantum spin Hall insulator, based on the Kane-Mele model defined in a finite zigzag graphene nanoribbon. We investigated the influence of the Kondo couplings with the helical liquids on both the static and dynamic properties of the ground state. The number and distinct spatial structures of the active natural orbitals (ANOs), which play essential roles in constructing the ground-state wave function, were first analyzed. Our numerical results indicate that two ANOs emerge, equal to the number of helical liquids. Specifically, at the coupling symmetry point, both ANOs are fully active with their spatial structures being respectively constituted by the different helical liquids. In comparison, when deviating from the symmetry point, only one ANO remains fully active, which is dominantly constructed by the helical liquid with the larger Kondo coupling. Local screening of the impurity, described by the impurity spin polarization and susceptibility, was further studied. It shows that at the coupling symmetry point, the impurity is maximally polarized and the spin susceptibility reaches the maximum. On the contrary, the impurity tends to be screened without polarization when the Kondo couplings deviate well from the symmetry point. The Kondo screening cloud, manifested by the spin correlation between the impurity and the conduction electrons, was finally explored. It is demonstrated that the Kondo cloud is mainly formed by the helical liquid with the larger Kondo coupling to the impurity. On the other hand, the spin-orbital coupling breaks the symmetry in spatial distribution of the spin correlation, leading to anisotropy in the Kondo cloud.
\end{abstract}

\maketitle

\section{Introduction}
\label{sec:introduction}
The Kondo effect\cite{Kondo1964,Hewson1997}, which results from the antiferromagnetic exchange interaction between a spin-$\frac{1}{2}$ impurity and conduction electrons, is one of the most intensively studied phenomena in strongly correlated many-body physics. Below a characteristic energy scale $T_K$ (the Kondo temperature), the impurity spin tends to be collectively screened by surrounding electrons, and finally it is perfectly screened at zero temperature, leading to a strong-coupling fixed point of Fermi liquid type in the renormalization group flow, and forming the Kondo screening cloud with a spatial extension determined by the Kondo length $\xi_K={\hbar v_F}/{k_B T_K}$ with $v_F$ the Fermi velocity\cite{Affleck1996,Affleck2001,Affleck2005,Bergmann2007,Holzner2009,Busser2010}. Over the past years, the Kondo screening cloud has been intensively investigated theoretically\cite{Gubernatis1987, Barzykin1996, Simon2002, Simon2003, Hand2006, Borda2007, Affleck2008, Pereira2008, Andrew2011, Yoshii2011, Jinhong2013, Busser2013, Shirakawa2014, Shirakawa2016} and experimentally\cite{Boyce1974,Madhavan1998,Manoharan2000,Y2007,Wenderoth2011,Borzenets2020}. Moreover, in a recent work\cite{Borzenets2020}, the experimental evidence of a Kondo cloud extending over a length of micrometres, comparable to the theoretical length $\xi_K$, was presented.

Furthermore, when two equivalent but independent channels of electrons compete to screen a spin-$\frac{1}{2}$ impurity, the strong quantum fluctuations occur. Ultimately, the impurity is overscreened at low temperatures, resulting in an intermediate-coupling fixed point\cite{Blandin1980} of non-Fermi liquid type, namely, the two-channel Kondo (2CK) effect emerging. However, at zero temperature, when the channel symmetry is broken, the impurity is finally screened by the electron channel with the larger Kondo coupling while the other channel decouples\cite{Blandin1980,Pang1991,Andrei1995,Mitchell2011,Mitchell2014}, leading to the standard single-impurity Kondo physics with a Fermi-liquid phase. As a consequence, the system may undergo a quantum phase transition driven by the channel asymmetry at zero temperature. Experimental observations of the 2CK and three-channel Kondo effect were also obtained\cite{Cichorek2005,Potok2007, Mebrahtu2013,Keller2015,Iftikhar2015,Zhu2015,Cichorek2016,Iftikhar2018}. Recently, the Kondo effect involved in the topological systems, including the single-impurity Kondo effect and the 2CK effect, has attracted extensive concerns.

The quantum spin Hall insulator (QSHI)\cite{Kane20052,Bernevig2006,Konig2007}, in which the spin-orbital coupling (SOC) plays an essential role\cite{Hasan2010,Qi2011}, is a kind of topologically nontrivial matter state. It possesses a finite bulk gap, but supports 1D gapless edge states with quantized conductance of $G = 2e^2/h$, which are called helical liquids\cite{CWu2006} due to that the Kramers' pair of states with opposite spin polarizations counterpropagate at each open edge. The helical edge states are robust against weak interactions and perturbations preserving the time-reversal symmetry (TRS)\cite{Kane20051,CWu2006,Xu2006}, since the TRS protects the edge electrons from backscattering. However, when a quantum impurity interacts with the helical edge states, the backscattering with spin-flip scattering is allowed, giving rise to the Kondo effect in the QSHIs, which may exhibit new features distinct from that in the normal metals. The single-impurity Kondo effect in the QSHIs has been investigated\cite{CWu2006,Maciejko2009,Yoichi2011,Joseph2012,Erik2013,Assaad2013,Hu2013}. It is shown that the helical edge states restore the perfect quantization of conductance $G$ at zero temperature\cite{Maciejko2009}, resulting from the complete screening of the impurity spin. Meanwhile, the SOC has substantial influence on the Kondo effect in the QSHIs\cite{Zitko2011, Zarea2012,Isaev2012,Kikoin2012,Grap2012,Mastrogiuseppe2014,Wong2016,Sousa2016}.

It is also surely intriguing to study the problem of a quantum impurity interacting with two helical liquids, where novel 2CK effect may occur. Based on a simplified model of the helical edge states, a quantum impurity coupled to two helical liquids has been studied by the Abelian bosonization technique\cite{Ng2010,Posske2013,Lee2013,Trauzettel2014}. If the electrons in the helical liquids are noninteracting with the Luttinger liquid parameter $K=1$, the system can be described by a one-channel Kondo (1CK) Hamiltonian\cite{Ng1988,Posske2013}. On the other hand, a weak repulsive interaction, imposed on the electrons at the edges with $K < 1$, is enough to drive the system to the 2CK fixed point at low temperatures\cite{Ng2010,Lee2013}. Nevertheless, the studies suffer from a lack of reliable numerical calculations on such systems, since conventional quantum many-body numerical methods encounter difficulties in solving such systems.

Graphene nanoribbon (GNR), a quasi one-dimensional material, has been intensively studied in recent years. It is generally accepted that GNRs exhibit extraordinary mechanical and electronic properties, and thus hold a promise for use in nanoelectronics\cite{Novoselov2004,Son2006,Han2007,Barbaros2007,Geim2007} and nanospintronics\cite{Igor2004,Louie2006,Oded2008,Han2014}. Advanced nanoscale technologies have also stimulated the fabrication of graphene-based devices, which could promote the applications of spintronics as well as the development of quantum computing. Helical liquids, which feature the novel spin-momentum locking property, appear in both armchair and zigzag GNRs with SOC\cite{Kane20052}. Therefore, it is interesting to investigate the Kondo effect in nanostructures such as topological GNRs. In addition, in a topological GNR with finite size, the helical edge states can come back by traversing the whole edge of the system, which may lead to distinct behaviors of the Kondo effect\cite{Sonu2019,Zheng20212}, especially that in the system with a spin-$\frac{1}{2}$ impurity coupled to two helical liquids.

In this work, utilizing the natural orbitals renormalization group (NORG) method\cite{He2014}, we investigated a localized spin-$\frac{1}{2}$ impurity coupled to two helical liquids, based on the Kane-Mele (KM) model\cite{Kane20052} defined in a zigzag graphene nanoribbon (ZGNR) with finite size. We explored how the Kondo couplings with the helical liquids influence the static and dynamic properties of the ground state.

The structure of this paper is organized as follows. In Sec.~\ref{sec:Model-Method} the model Hamiltonian as well as the NORG method are introduced. The energy spectrum of the KM model is first presented in Sec.~\ref{sec:EnergySpetrum}. In Sec.~\ref{sec:ANOs} we investigate the number and spatial structures of the active natural orbitals (ANOs) for the many-body ground state, which are helpful for clarifying the intrinsic structure of the ground state. In the following Sec.~\ref{sec:moment}, we further study the local screening of the impurity spin, described by the spin polarization and local susceptibility at the impurity site. Finally, in Sec.~\ref{sec:ScreeningCloud}, we explore the Kondo screening cloud, which is represented by the spin correlation between the impurity and the conduction electrons. Section~\ref{sec:summary} gives a summary of this work.

\section{Model and numerical method}
\label{sec:Model-Method}
\begin{figure}[htbp!]
\centering
\includegraphics[width=0.7\columnwidth]{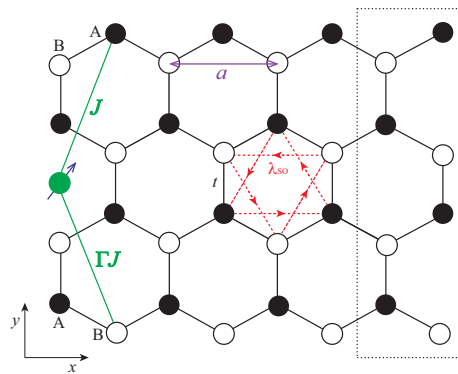}
\caption{\label{fig:Model}(color online) Schematic of the system. A local spin-$\frac{1}{2}$ impurity is coupled to two helical liquids localized at two open edges in a QSHI, simulated by the ground state of the KM model defined in a ZGNR. The unit cell of the ZGNR is shown as the dotted black rectangle. The ZGNR is periodic (open) along the $x$ $(y)$ direction with length $N_x=4$ (width $N_y=4$). The width $N_y$ is defined by the number of zigzag lines. Sublattice A (B) is denoted by the black filled (open) circles. The black lines represent the NN hopping connecting two different sublattices. The NNN hopping connecting the same sublattice, namely the SOC term, is denoted as the red dashed arrows with signs associated with $\nu_{ij}$. The local spin-$\frac{1}{2}$ impurity, marked by the filled green circle, is coupled to the helical liquid along the top (bottom) edge via the Kondo coupling $J_t=J$ ($J_b = \Gamma J$). Here $a$ stands for the lattice constant and $J>0$ the Kondo coupling strength.}
\end{figure}

The KM model is a prototypical framework to study the helical edge states in the QSHIs. We consider the following Hamiltonian $H_{\text {KM}}$ of the KM model,
\begin{equation}
H_{\text {KM}} = -t\sum\limits_{\langle ij\rangle\sigma}c_{i\sigma }^ {\dagger}{c_{j\sigma }} +
              i\lambda_{\text{SO}}\sum\limits_{\langle\langle ij\rangle\rangle \alpha\beta}\nu_{ij}c_{i\alpha}^{\dagger}\sigma_{\alpha\beta}^zc_{j\beta}.
\label{eq:KM}
\end{equation}
Here $c_{i\sigma}$ ($c_{i\sigma}^{\dagger}$) denotes the annihilation (creation) operator of a conduction electron at the $i$th site with spin $\sigma=\uparrow,\downarrow$. The first term represents the nearest-neighbor (NN) hopping with an amplitude of $t$. The second term is an intrinsic SOC involving the next-nearest-neighbor (NNN) hopping with an imaginary hopping integral of $\lambda_{\text{SO}}$. The parameter $\nu_{ij}$ is determined by the orientation of two NN bonds that an electron hops from site $j$ to $i$, specifically $\nu_{ij} = +1$ ($-1$) if the electron turns left (right) in the hopping from site $j$ to $i$, as depicted in Fig.~\ref{fig:Model}. The Pauli matrix $\sigma_{\alpha\beta}^z$ further distinguishes the spin-up and spin-down edge states with opposite NNN hopping amplitudes.

For the ground state of the KM model [Eq.~(\ref{eq:KM})] defined in a ZGNR, two helical edge states are exponentially localized at each open edge, corresponding to the noninteracting limit of helical Luttinger liquid. For simplicity, we accordingly set a local spin-$\frac{1}{2}$ impurity coupled to both open edges in the ZGNR, namely, we construct a 2CK model with helical Luttinger liquid. Since the edge states mainly reside in sublattice A (B) at the top (bottom) edge\cite{Andrew2017b,Zheng2018,Cazalilla2018}, the impurity is coupled to an edge site of sublattice A (B) at the top (bottom) edge, as illustrated in Fig.~\ref{fig:Model}. Then the total Hamiltonian $H$ of the system is given by the following forms,
\begin{equation}
\begin{array}{l}
H = H_{\text {KM}} + H_{\text {Kondo}}, \\
H_{\text {Kondo}}=J{\bf{S}}_{\text {imp}} \cdot {\bf{s}}_{t0} + \Gamma J {\bf{S}}_{\text {imp}} \cdot {\bf{s}}_{b0},
\end{array}
\label{eq:Model}
\end{equation}
where $H_{\text{Kondo}}$ describes the antiferromagnetic exchange interaction between the impurity and edge-state electrons. The local spin $\bf{S}_{\text {imp}}$ is coupled to the electron spin ${\bf{s}}_{t(b)0}=\frac{1}{2}\sum_{\alpha\beta}c_{0\alpha}^ {\dagger}{{\bf{\sigma}}_{\alpha\beta}}{c_{0\beta}}$ at the $0$th site along the top (bottom) edge via the Kondo coupling $J_t = J$ ($J_b = \Gamma J$) with ${\bf{\sigma}}$ denoting the vector of Pauli matrices. Here the dimensionless quantity $\Gamma$ plays a controlling role, which governs the coupling symmetry between the two helical liquids with the impurity. In the case of $\Gamma =0$ with $J_b=0$, the impurity only interacts with the conduction electrons at the top edge, and it is perfectly screened by the top helical liquid at zero temperature\cite{Maciejko2009}. For any finite $\Gamma \neq 0$, the impurity interacts with both helical liquids, and it is completely screened by the two helical liquids together\cite{Posske2013}. Specifically, at the point of $\Gamma =1$ with $J_t=J_b$, i.e., at the coupling symmetry point, the two helical liquids equivalently screen the impurity spin. However, when the controlling parameter $\Gamma \neq 1$, the coupling symmetry between the two helical liquids with the impurity is broken with $J_t \neq J_b$. In consequence, at zero temperature, screening of the impurity by each helical liquid may vary as the Kondo couplings deviate from the symmetry point. We thus focus on the behavior of the spin-$\frac{1}{2}$ impurity as well as that of the Kondo screening cloud around the coupling symmetry point.

On the other hand, for a conventional 2CK model, the Hamiltonian obeys exact ${\rm SU}(2)_{\rm spin}$ symmetry. However, if the ${\rm SU}(2)_{\rm spin}$ symmetry is broken, the system will be driven to flow to a polarized Fermi-liquid fixed point\cite{Pang1991}, indicating that the 1CK physics emerges. In comparison, as we see, the total Hamiltonian $H$ [Eq.~(\ref{eq:Model})] preserves ${\rm U}(1)_{\rm spin}$ symmetry with $[S_{\rm total}^z, H] = 0$, since the ${\rm SU}(2)_{\rm spin}$ symmetry is partially broken by the SOC. Hence in this sense, we expect that properties of the 2CK model with two helical liquids in our work should exhibit the 1CK physics, as indicated in Refs.~\onlinecite{Ng1988,Posske2013} that such a system can be described by the usual 1CK Hamiltonian for non-interacting electrons on the edges with the Luttinger liquid parameter $K=1$.

We employed the NORG approach, a many-body numerical method without perturbation, to study the 2CK model with two helical liquids, as depicted in Fig.~\ref{fig:Model} and Eq.~(\ref{eq:Model}). It has been demonstrated that the NORG is an effective method to study quantum impurity problems\cite{He2014,He2015,Zheng2018,Zheng2020,Zheng2021,Zheng20212}. Generally, the NORG method preserves the whole geometric information of a lattice, and its effectiveness is independent of any lattice structure or topology of a quantum impurity system.

For a lattice system with finite size $L$, the NORG involves a representation transformation from the site representation into the natural orbital representation by $d_{m}^{\dagger}=\sum_{i=1}^LV_{mi}^{\dagger}c_{i}^{\dagger}$ with $d_{m}^{\dagger}$ ($c_{i}^{\dagger}$) representing the creation operator in the natural orbital (site) representation and $V$ an $L \times L$ unitary matrix. Thus the NORG method works in the Hilbert space constructed from a set of NOs, which correspond to the eigenvectors of the single-particle density matrix (or the correlation matrix)~\cite{PerOlov1955,Luo2010,Zgid2012,Lin2013,Lu2014,He2014,Fishman2015,Lu2019} defined by $D_{ij}  = \langle\Psi| c_{i}^{\dagger}c_{j}|\Psi\rangle$ with $|\Psi\rangle$ a normalized many-body wave function of the system. Thus the single-particle density matrix $D$ is diagonalized by the unitary matrix $V$. On the other hand, the associated eigenvalues of $D$ correspond to the occupancy numbers of the NOs. A natural orbital is active if its occupancy number deviates well from empty or full occupancy, otherwise inactive. In a quantum impurity system, it has been shown that\cite{He2014,Zheng2018,Zheng2020} the ANOs play substantial roles in constructing the ground-state wave function while the inactive NOs are frozen as background, and the number of ANOs is roughly equal to that of the interacting impurities. This is also the underlying basis for the NORG method working on quantum impurity problems. In our practical calculations, to efficiently realize the NORG approach, we only rotate the bath orbitals, i.e., only the bath sites are transformed into the NOs representation. As a consequence, the computational cost needed is about $O(L^3)$ and thus we can solve hundreds of noninteracting bath sites.

In the following calculations, we set the NN hopping parameter $t=1.0$ as the unit of energy and keep half-filling for the conduction band. The strength of SOC is set to $\lambda_{\text{SO}}=0.1t$, a value appropriate for the germanene or stanene, and that of antiferromagnetic Kondo coupling is set to $J=t$. All the calculations were performed in the ground-state subspace of $S_{\rm total}^z=\frac{1}{2}$. The system size is $L = N_x\times N_y$ with $N_x$ ($N_y$) denoting the length (width) of the ZGNR, and the periodic (open) boundary condition was adopted along the $x$ $(y)$ direction, as schematically shown in Fig.~\ref{fig:Model}.

\section{Numerical results}
\label{sec:Results}
\subsection{Band structure}
\label{sec:EnergySpetrum}
\begin{figure}[htp!]
\centering
\includegraphics[width=1.0\columnwidth]{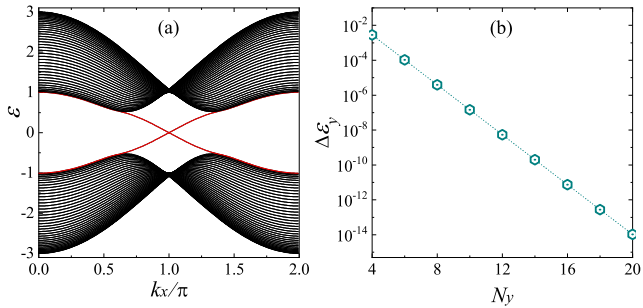}
\caption{\label{fig:EnergySpectra}(color online) (a) Energy spectrum of the KM model in a ZGNR and (b) the finite-size energy gap $\Delta \varepsilon_y$ at the Fermi wave vector $k_F = \pi$ for the SOC strength $\lambda_{\text{SO}} = 0.1$, respectively. The edge-state bands [marked by the red lines in (a)] cross with each other at the Fermi level $\varepsilon_F = 0$, and each band is doubly degenerate according to the Kramers degeneracy. On the other hand, the energy gap $\Delta \varepsilon_y$ decays exponentially as the width $N_y$ increases in finite-size systems. The length (width) of the ZGNR adopted in the calculations in (a) is $N_x=256$ ($N_y=40$). The ZGNR length is fixed to be $N_x = 28$ in (b).}
\end{figure}

The energy spectrum of the KM model in a ZGNR was first calculated, as plotted in Fig.~\ref{fig:EnergySpectra}(a). As we see, two bands of the localized helical edge states cross with each other at the Fermi level $\varepsilon_F= 0$, and each band is doubly degenerate according to the Kramers degeneracy. Practically, in a realistic system, the localized edge states have finite width and decay exponentially into the bulk\cite{Niu2008}. As a result, in a ZGNR of finite width, the helical edge states coming from different edges can couple together with a finite overlap to generate a small energy gap at the Fermi wave vector. Figure~\ref{fig:EnergySpectra}(b) shows the finite-size gap $\Delta \varepsilon_y$ at the Fermi wave vector $k_F = \pi$ with respect to the ZGNR width $N_y$. As expected, the energy gap $\Delta \varepsilon_y$ decays exponentially with the width $N_y$ increasing. In the following calculations, we always keep the width of ZGNRs to be $N_y=10$ without additional statement with the corresponding finite-size energy gap $\Delta \varepsilon_y \approx 10^{-7}$, appropriate to simulate the localized helical edge states.

\subsection{Active natural orbitals}
\label{sec:ANOs}
In this section, we analyze the NOs and their occupancy numbers. For a quantum impurity system, the ANOs, whose occupancy numbers deviate well from empty or full occupancy, play essential roles in constructing the ground-state wave function while the inactive NOs are frozen as background\cite{He2014,Zheng2018,Zheng2020}. We thus focus on the number as well as structures of the ANOs for the ground state, which are helpful for clarifying the intrinsic structure of the many-body ground state with strong correlation and high entanglement in such systems.

\begin{figure}[htp!]
\centering
\includegraphics[width=1.0\columnwidth]{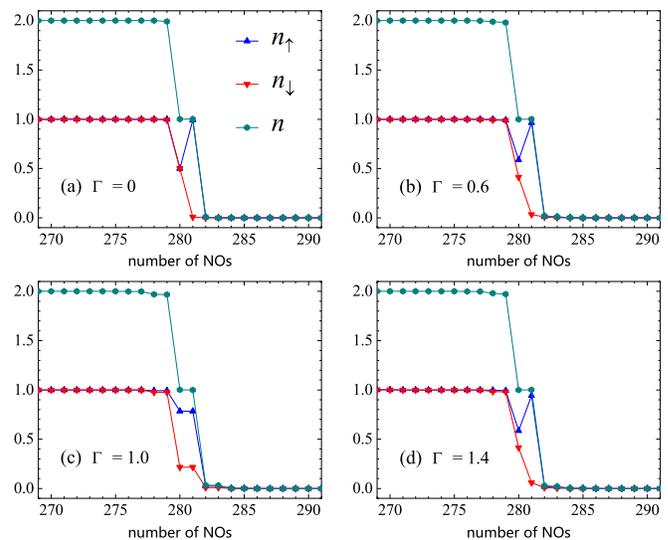}
\caption{\label{fig:OccNOs}(color online) The total occupancy number $n$ ($0\leq n \leq 2.0$) and the corresponding component $n_\sigma$ ($0\leq n_\sigma \leq 1.0$) of the NOs for the ground state with different values of the controlling parameter $\Gamma$. Here most of the NOs exponentially decay into full occupancy with $n=2.0$ or empty with $n=0$, except two ANOs of half-occupied with total occupancy number $n = 1.0$. In comparison, the component $n_\sigma$ of the two ANOs depends on the values of $\Gamma$. The calculations were carried out in the ZGNR of length $N_x=28$.}
\end{figure}

Figure.~\ref{fig:OccNOs} shows the total occupancy number $n$ ($0\leq n \leq 2.0$) of the NOs for different values of $\Gamma$. As expected, most of the NOs exponentially rush into full occupancy with $n=2.0$ or empty with $n=0$, indicating that they are inactive NOs. In comparison, there are two NOs half-occupied with total occupancy number $n=1.0$, which deviates well from empty or full occupancy, suggesting that there are essentially two ANOs for the ground state. Then we further examine the corresponding component $n_\sigma$ ($0\leq n_\sigma \leq 1.0$) of the two ANOs with $n=n_\uparrow + n_\downarrow$, which is expected to depend on the coupling symmetry between the two helical liquids with the impurity, governed by the values of the parameter $\Gamma$.

More specifically, as presented in Fig.~\ref{fig:OccNOs}(c), at the coupling symmetry point of $\Gamma =1$ with $J_t=J_b$, the components $n_{\uparrow}$ and $n_{\downarrow}$ of the first ANO (indexed as the $280$th NO in Fig.~\ref{fig:OccNOs}) are respectively equal to those of the second ANO (indexed as the $281$th NO in Fig.~\ref{fig:OccNOs}). In contrast, in the case of $\Gamma=0$ with $J_b=0$, $n_{\uparrow}=n_{\downarrow}=0.5$ for the first ANO, while $n_{\uparrow}\simeq 1.0, n_{\downarrow}\simeq 0$ for the other, as plotted in Fig.~\ref{fig:OccNOs}(a). Likewise, when $\Gamma$ is sufficiently large with $J_t << J_b$ (such as $\Gamma=4$, not plotted), our numerical results show that $n_{\uparrow} \simeq 0.5, n_{\downarrow} \simeq 0.5$ for the first ANO, while $n_{\uparrow}\simeq 1.0,n_{\downarrow}\simeq 0$ for the other. Hence when $\Gamma$ goes to small or large values from the symmetry point $\Gamma=1$, the corresponding component $n_{\uparrow}$ ($n_{\downarrow}$) of the first ANO tends to decrease (increase) to $0.5$, while that of the other ANO tends to increase (decrease) to $1.0$ ($0$), as shown in Figs.~\ref{fig:OccNOs}(b) and ~\ref{fig:OccNOs}(d), respectively. As a result, when the Kondo couplings with the two helical liquids deviate from the symmetry point, the first ANO remains active, while the other tends to be inactive.

In general, we identify an NO with total occupancy number $n=n_\uparrow + n_\downarrow$ deviating well from full occupancy or empty to be an ANO, regardless of the values of their components $n_\uparrow$ and $n_\downarrow$. Here we further consider an ANO with its components $n_\uparrow$ and $n_\downarrow$ both deviating well from full occupancy ($n_\sigma=1.0$) or empty ($n_\sigma=0$) to be fully active. In this sense, the first ANO is identified as a full ANO in the whole regimes of $\Gamma$. Hence we come to the conclusion that in a quantum impurity system of a local spin-$\frac{1}{2}$ impurity interacting with two helical liquids, the number of ANOs for the ground state is equal to that of the helical liquids. All the ANOs become fully active when the Kondo couplings keep symmetric, while only one remains fully active when deviating from the coupling symmetry point.

\begin{figure}[htp!]
\centering
\includegraphics[width=1.0\columnwidth]{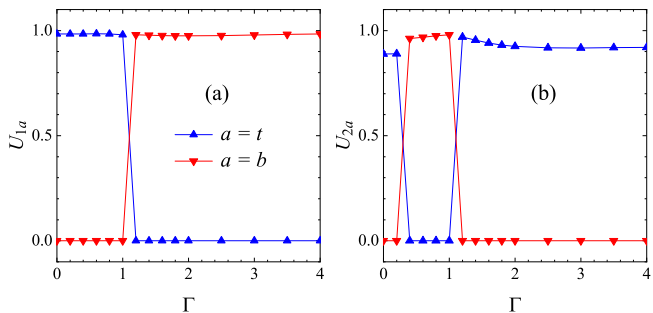}
\caption{\label{fig:StructuresANOs}(color online) Amplitudes of (a) the first and (b) the second ANOs projected into the real space along the top and bottom edges, respectively. Spatial structures of the two ANOs with respect to the parameter $\Gamma$ are distinct and depend on the value of $\Gamma$. Our numerical results were obtained in the ZGNR of length $N_x=28$.}
\end{figure}

In order to clarify the structures of the ANOs, we project them into the real space along the top and bottom edges in the ZGNR. From the representation transformation $d_{m\sigma}^{\dagger}=\sum_{i=1}^NV_{mi\sigma}^{\dagger}c_{i\sigma}^{\dagger}$ involved in the realization of the NORG, we obtain the amplitude $U_{ma}$ of the $m$th ANO projected into the $a \in \left\{t,b\right\}$ edge by $U_{ma\sigma}=\sum_{i\in a}|V_{mi\sigma}^ {\dagger}|^2$. Our results (not plotted) show that $U_{ma\uparrow} = U_{ma\downarrow}$, leading to $U_{ma} = U_{ma\sigma}$, meaning that the amplitude $U_{ma}$ is spin independent. Our results also show that the spatial structures of the two ANOs are distinct and depend on the numerical value of $\Gamma$, as presented in Fig.~\ref{fig:StructuresANOs}.

Even though the first ANO remains fully active in the whole regimes of the controlling parameter $\Gamma$, its spatial structure varies around the coupling symmetry point. From Fig.~\ref{fig:StructuresANOs}(a), we find that $U_{1t} \simeq 1$ while $U_{1b} \simeq 0$ for $\Gamma \leq 1$, meaning that it is mainly constructed by the orbitals at the top edge. In contrast, for any value of $\Gamma > 1$, $U_{1b} \simeq 1$ while $U_{1t} \simeq 0$, indicating that it is dominantly composed by the orbitals at the bottom edge. As a result, in the case of $J_t > J_b$ ($J_t<J_b$), the first ANO is constituted by the helical liquid at the top (bottom) edge, namely, the helical liquid with the larger Kondo coupling plays a dominant role. Moreover, at the coupling symmetric point of $J_t=J_b$, the components of the first ANO come from the helical liquid at the top edge.

In comparison, as shown in Fig.~\ref{fig:StructuresANOs}(b), the spatial structure of the second ANO is distinct from that of the first ANO, especially in the regime of $\Gamma < 1$ with $J_t > J_b$. As we see, for small $\Gamma$ (for example $\Gamma = 0.2$) with $J_t >> J_b$, the helical liquid at the top edge tends to form the second ANO with $U_{2t} \to 1$ while $U_{2b} \simeq 0$. As $\Gamma$ further increases (for example $0.4 \leq \Gamma \leq 1$), $U_{2b}$ goes to $1$ while $U_{2t}$ tends to vanish, illustrating that the helical liquid at the bottom edge dominantly participates in constituting the ANO. Furthermore, in the case of $\Gamma > 1$ with $J_t < J_b$, it is mainly comprised by the helical liquid at the top edge with $U_{2t} \to 1$ while $U_{2b} \simeq 0$. Compared with the first ANO, at the coupling symmetry point of $\Gamma=1$, the second ANO is formed by the helical liquid at the bottom edge. Consequently, for the ground state at the coupling symmetry point, two full ANOs emerge with one being constituted by the helical liquid at the top edge while the other by that at the bottom edge.

\subsection{Impurity spin polarization and susceptibility}
\label{sec:moment}
\begin{figure}[htp!]
\centering
\includegraphics[width=1.0\columnwidth]{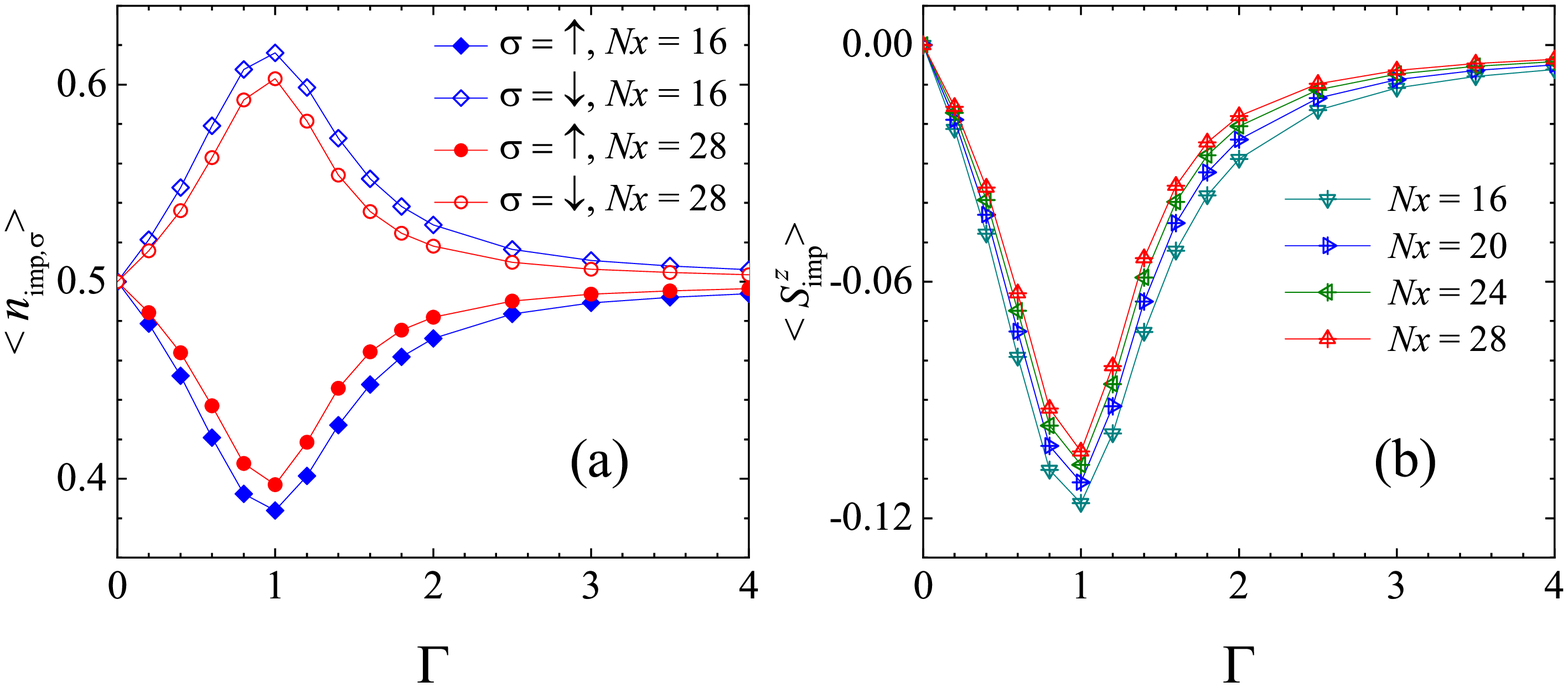}
\caption{\label{fig:Simpz}(color online) (a) The occupancy number $\langle n_{\text {imp}, \sigma}\rangle$ and (b) the spin polarization $\langle S_{\text {imp}}^z\rangle=\frac{1}{2}(\langle n_{\text {imp}, \uparrow}\rangle-\langle n_{\text {imp}, \downarrow}\rangle)$ at the impurity site for the ground state, respectively. At the coupling symmetry point of $\Gamma = 1$, the occupancy number $\langle n_{\text {imp}, \downarrow}\rangle$ ($\langle n_{\text {imp}, \uparrow}\rangle$) reaches a maximum (minimum). Correspondingly, the impurity spin polarization $\langle S_{\text {imp}}^z\rangle$ approaches to the maximal value with a negative sign at $\Gamma = 1$. The local spin is negatively polarized in the ground-state subspace of $S_{\rm total}^z=1/2$ for any finite $\Gamma \neq 0$, and it is maximally polarized at the coupling symmetry point. Furthermore, the impurity spin polarization $\langle S_{\text {imp}}^z\rangle$ decreases with the length $N_x$ of the ZGNRs increasing.}
\end{figure}

To proceed, we study the local screening of the impurity, which is described by the impurity spin polarization and susceptibility. We first investigate behavior of the impurity spin polarization, defined as $\langle S_{\text {imp}}^z\rangle=\frac{1}{2}(\langle n_{\text {imp}, \uparrow}\rangle-\langle n_{\text {imp}, \downarrow}\rangle$) with $\langle n_{\text {imp}, \sigma}\rangle$ representing the occupancy number with spin $\sigma$ at the impurity site. Figures~\ref{fig:Simpz}(a) and ~\ref{fig:Simpz}(b) present the numerical results of the occupancy number $\langle n_{\text {imp}, \sigma}\rangle$ and the spin polarization $\langle S_{\text {imp}}^z\rangle$ for the ground state, respectively.

As shown in Fig.~\ref{fig:Simpz}(a), at the point of $\Gamma=0$ with $J_b = 0$, the occupancy numbers $\langle n_{\text {imp}, \uparrow}\rangle=\langle n_{\text {imp}, \downarrow}\rangle=0.5$. As $\Gamma$ further increases, $\langle n_{\text {imp}, \downarrow}\rangle$ ($\langle n_{\text {imp}, \uparrow}\rangle$) increases (decreases) to a maximum (minimum) at the coupling symmetry point of $\Gamma = 1$, while $\langle n_{\text {imp}, \downarrow}\rangle$ ($\langle n_{\text {imp}, \uparrow}\rangle$) tends to decrease (increase) to $0.5$ as $\Gamma$ increases to large values afterwards. Correspondingly, as shown in Fig.~\ref{fig:Simpz}(b), the impurity spin polarization $\langle S_{\text {imp}}^z\rangle$ reaches the maximum with a negative sign at $\Gamma = 1$, and then it begins to decay as $\Gamma$ goes to rather small or large values. Consequently, the impurity spin is negatively polarized in the ground-state subspace of $S_{\rm total}^z=1/2$ for any finite $\Gamma \neq 0$, and its quantity manifests the maximum at the coupling symmetry point of $\Gamma=1$. Moreover, $\langle S_{\text {imp}}^z\rangle$ tends to vanish in the regime of $\Gamma << 1$ or $\Gamma >> 1$, meaning that the impurity spin tends to be completely screened without polarization in the case of $J_t >> J_b$ or $J_t << J_b$. Hence in a finite system, the impurity spin is maximally polarized at the coupling symmetry point, while its polarization is suppressed if the symmetry of the Kondo couplings with the two helical liquids is broken.

In addition, we find that the impurity spin polarization decreases with the length $N_x$ of the ZGNRs increasing, as shown in Fig.~\ref{fig:Simpz}(b), meaning that $\langle S_{\text {imp}}^z\rangle$ may approach to vanish in the thermodynamic limit $1/N_x \to 0$ for any value of $\Gamma$. This implies that the local spin tends to be screened without free local moment formed at the impurity site in the thermodynamic limit, i.e., the standard 1CK physics emerges, which is consistent with the previous conclusion that such a system can be described by the usual 1CK Hamiltonian for non-interacting electrons on the edges\cite{Ng1988,Posske2013}.

\begin{figure}[htp!]
\centering
\includegraphics[width=0.8\columnwidth]{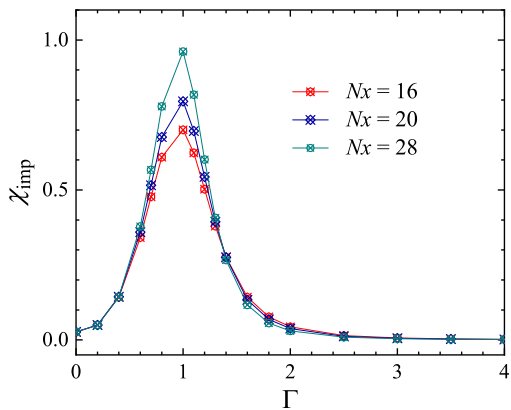}
\caption{\label{fig:Susceptibility}(color online) The impurity spin susceptibility $\chi_{\text {imp}}$ with the broadening factor $\eta=t/L$ and $L=N_x\times N_y$. The behavior of $\chi_{\text {imp}}$ is consistent with that of the impurity spin polarization $\langle S_{\text {imp}}^z\rangle$. The susceptibility $\chi_{\text {imp}}$ reaches the maximum at the coupling symmetry point of $\Gamma=1$ and then decays when $\Gamma$ deviates from this symmetry point. In the regime of $\Gamma << 1$ or $\Gamma >> 1$, the susceptibility $\chi_{\text {imp}}$ tends to vanish, meaning that the impurity is screened.}
\end{figure}

The impurity spin susceptibility, representing dynamic property of the system, is further calculated. We consider the following zero-temperature Green's function $G_{\text {imp}}$ defined at the impurity site,
\begin{equation}
G_{\text {imp}}(\omega) = \langle0|S_{\text {imp}}^z\frac{1}{\omega+i\eta-H+E_0}S_{\text {imp}}^z|0\rangle,
\label{eq:Greenfunction}
\end{equation}
where $|0\rangle$ and $E_0$ represent the ground state and ground-state energy, respectively. The parameter $\eta \to 0$ denotes a Lorentzian broadening factor. The above Green's function $G_{\text {imp}}$ for a given frequency $\omega$ is calculated using the correction vector method\cite{Kuhner1999,Jeckelmann2002} in our following calculations. Then the spin-excitation spectrum $A_{\text {imp}}(\omega)$ at the impurity site is obtained by $A_{\text {imp}}(\omega)=-\frac{1}{\pi}{\text{Im}}G_{\text {imp}}(\omega)$, and the impurity spin susceptibility $\chi_{\text {imp}}$ is further obtained by $\chi_{\text {imp}}=A_{\text {imp}}(\omega = 0)$. Numerical results of the susceptibility $\chi_{\text {imp}}$ are plotted in Fig.~\ref{fig:Susceptibility}.

As we see in Fig.~\ref{fig:Susceptibility}, the impurity spin susceptibility $\chi_{\text {imp}}$ changes nonmonotonic with $\Gamma$, similar to the behavior of the impurity spin polarization $\langle S_{\text {imp}}^z\rangle$. The susceptibility $\chi_{\text {imp}}$ reaches the maximum at the point of $\Gamma=1$ and then decays when $\Gamma$ deviates from the coupling symmetry point. Furthermore, $\chi_{\text {imp}}$ tends to vanish in the regime of $\Gamma << 1$ or $\Gamma >> 1$, meaning that the impurity spin is screened without local susceptibility in the case of $J_t >> J_b$ or $J_t << J_b$, consistent with the behavior of the impurity spin polarization $\langle S_{\text {imp}}^z\rangle$.

\subsection{Kondo screening cloud}
\label{sec:ScreeningCloud}
\begin{figure}[htp!]
\centering
\includegraphics[width=1.0\columnwidth]{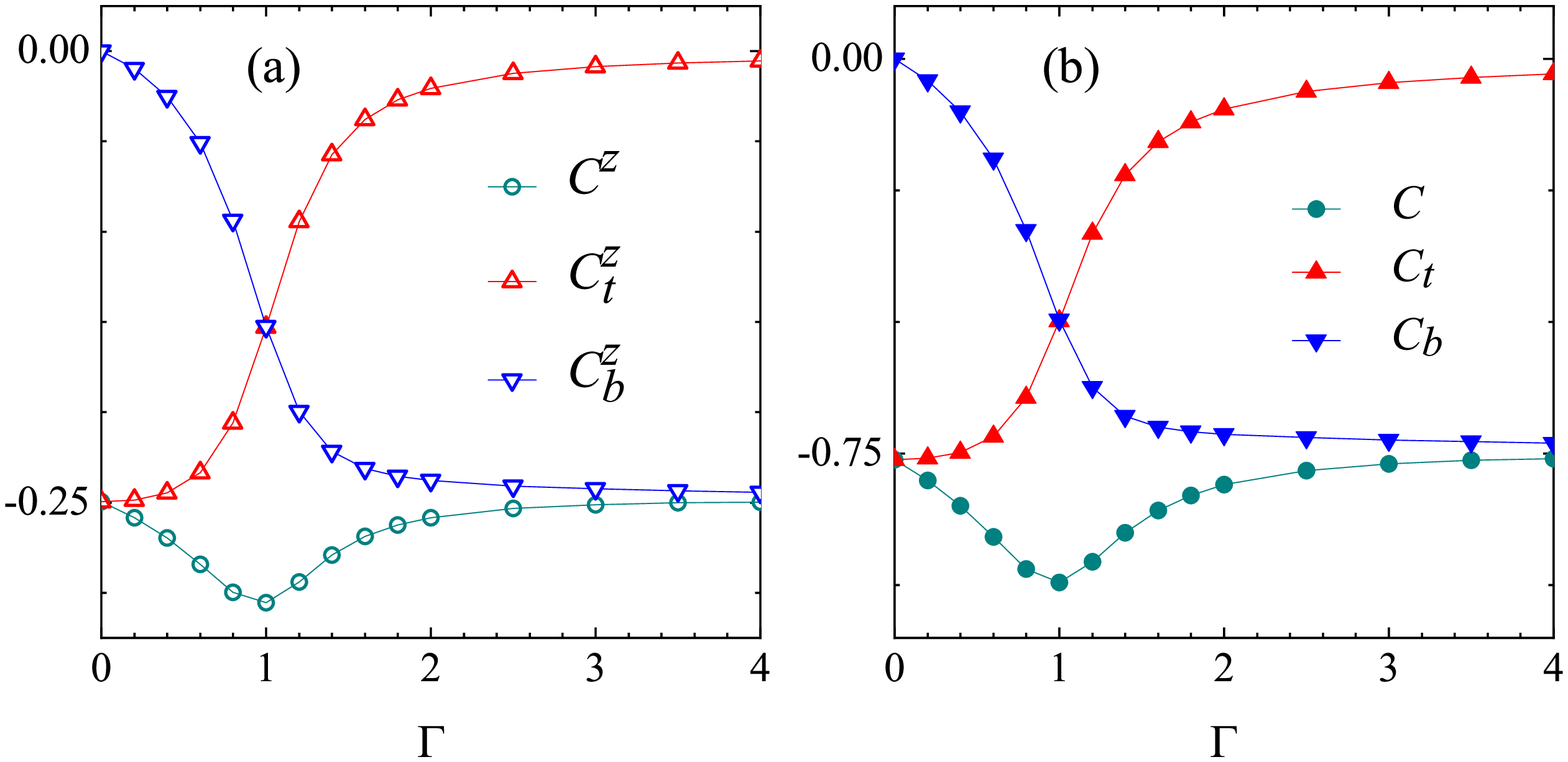}
\caption{\label{fig:Corr}(color online) (a) Spin correlations $C^z= C_t^z + C_b^z$ as well as $C_a^z$ with $a \in \left\{t,b\right\}$ between the impurity and the edge electrons projected into the $z$-direction. (b) Spin correlation $C_a$ between the impurity and the edge electrons at the $a \in \left\{t,b\right\}$ edge and the corresponding total correlation $C=C_t + C_b$. As $\Gamma$ increases from $0$ to large values, $C_t^z$ increases from $-\frac{1}{4}$ to $0$, while $C_b^z$ decreases from $0$ to $-\frac{1}{4}$. At the coupling symmetry pint of $\Gamma=1$, the spin correlations $C_t^z$ and $C_b^z$ cross, where the spin correlation $C^z$ reaches the minimal value accordingly. Behaviors of $C_a$ with $a \in \left\{t,b\right\}$ and $C$ are consistent with those of their components in the $z$-direction. The spin correlation functions were calculated in the ZGNR of length $N_x=28$.}
\end{figure}

Finally, we study the Kondo screening cloud, which is measured by the spin correlation between the impurity and the conduction electrons. We focus on the spin correlation functions in the ground state given by the following forms:
\begin{equation}
\begin{array}{l}
C_a^z = \sum\limits_{i \in a}\langle 0|S_{\text {imp}}^zs_{i}^z|0\rangle, \\
C_a = \sum\limits_{i \in a}\langle 0|{\bf{S}}_{\text {imp}} \cdot {\bf{s}}_{i}|0\rangle,
\end{array}
\label{eq:correlation}
\end{equation}
where $|0\rangle$ denotes the ground state. $C_a$ stands for the spin correlation between the impurity and the edge electrons at the $a \in \left\{t,b\right\}$ edge and $C_a^z$ represents the corresponding $z$-component. The total spin correlation $C=C_t + C_b$ between the impurity and the edge electrons at both open edges along with its component $C^z= C_t^z + C_b^z$ in the $z$-direction are analyzed as well. Numerical results of $C_a^z$, $C_a$, $C^z$, and $C$ are shown in Fig.~\ref{fig:Corr}.

From Fig.~\ref{fig:Corr}(a), at the point of $\Gamma=0$ with $J_b=0$, the spin correlations $C_t^z \simeq -\frac{1}{4}$ and $C_b^z \simeq 0$ as expected, meaning that the impurity is mainly correlated with the helical liquid at the top edge. As $\Gamma$ further increases, the correlation $C_t^z$ tends to increase from $-\frac{1}{4}$ to $0$. In contrast, $C_b^z$ decreases from $0$ to $-\frac{1}{4}$ with $\Gamma$ increasing. Hence, the impurity is correlated dominantly with the helical liquid with the larger Kondo coupling. At the coupling symmetry point of $\Gamma=1$ with $J_t=J_b$, the correlations $C_t^z$ and $C_b^z$ cross, indicating that the impurity is correlated equally with the two helical liquids. Accordingly, the $z$-component of total correlation $C^z= C_t^z + C_b^z$ decreases (increases) with $\Gamma$ in the regime of $\Gamma <1$ ($\Gamma >1$) and it approaches to a minimum at the coupling symmetry point.

As a comparison, as presented in Fig.~\ref{fig:Corr}(b), the behaviors of the spin correlations $C_a$ with $a \in \left\{t,b\right\}$ and $C=C_t + C_b$ are consistent with those of their components in the $z$-direction, respectively. In consequence, at the coupling symmetry point of $J_t=J_b$, the Kondo screening cloud is equally formed by the edge-state electrons in the two helical liquids, respectively. Furthermore, when the coupling symmetry breaks, the Kondo screening cloud is mainly formed by the electrons in the helical liquid with the larger Kondo coupling, i.e., the Kondo cloud is concentrated at the top (bottom) edge in the case of $J_t > J_b$ ($J_t < J_b$).

\begin{figure}[htp!]
\centering
\includegraphics[width=0.8\columnwidth]{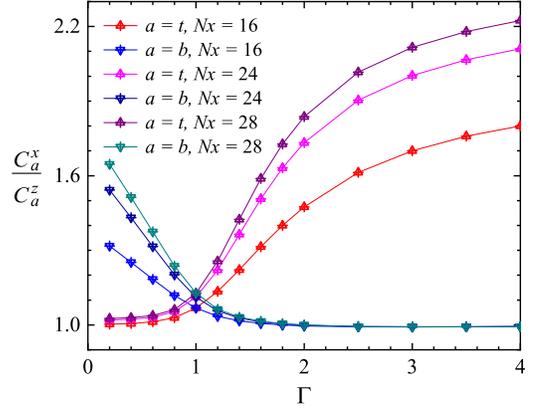}
\caption{\label{fig:Ratio}(color online) The ratio of $x$-component to $z$-component for the spin correlation $C_a$ with respect to the parameter $\Gamma$ with $a \in \left\{t,b\right\}$. The ratio of $\frac{C_a^x}{C_a^z}$ exhibits the influence of the SOC effect on the symmetry in spatial distribution of the Kondo cloud projected into the $a$ edge. Numerical results were calculated in the ZGNRs of length $N_x=16$, $N_x=24$, and $N_x=28$.}
\end{figure}

On the other hand, the SOC effect, which partially breaks the ${\rm SU}(2)_{\rm spin}$ symmetry of the system [Eq.~(\ref{eq:Model})], will influence the spatial distribution of the Kondo cloud. We hence explore the ratio of $x$-component to $z$-component for the spin correlation $C_a$ with respect to the parameter $\Gamma$, namely, the ratio of $\frac{C_a^x}{C_a^z}$ with $a \in \left\{t,b\right\}$. Numerical results are shown in Fig.~\ref{fig:Ratio}. We find that $\frac{C_t^x}{C_t^z} \neq 1$ and $\frac{C_b^x}{C_b^z} \neq 1$, demonstrating that the spatial symmetry of the spin correlation projected into the $a \in \left\{t,b\right\}$ edge is broken. More specifically, as the parameter $\Gamma$ increases, the ratio of $\frac{C_b^x}{C_b^z}$ decays, which is consistent with the behavior of the spin correlation $C_b$, while $\frac{C_t^x}{C_t^z}$ increases. Therefore, the Kondo cloud in the system is anisotropic, which is in contrast to the isotropic Kondo cloud formed in a normal metal.

Furthermore, the ratio of $\frac{C_t^x}{C_t^z} \to 1$ ($\frac{C_b^x}{C_b^z} \to 1$) in the regime of $\Gamma <<1$ ($\Gamma >>1$), meaning that the symmetry in spatial distribution of the Kondo cloud projected into the top (bottom) edge tends to recover in the small (large) $\Gamma$ regime. Since the Kondo cloud is dominantly concentrated at the top (bottom) edge in the case of $\Gamma <1$ ($\Gamma >1$), and that $\frac{C_t^x}{C_t^z} << \frac{C_b^x}{C_b^z}$ ($\frac{C_t^x}{C_t^z} >> \frac{C_b^x}{C_b^z}$) for $\Gamma <1$ ($\Gamma >1$) shown in Fig.~\ref{fig:Ratio}, the spatial symmetry of the Kondo cloud that mainly screens the local impurity is weakly influenced by the SOC effect. In addition, the values of $\frac{C_t^x}{C_t^z}$ and $\frac{C_b^x}{C_b^z}$ increase with the length $N_x$, as shown in Fig.~\ref{fig:Ratio}, indicating that the influence of the SOC term on the Kondo cloud is enhanced by the size of the ZGNRs.

\section{summary}
\label{sec:summary}
To summarize, by means of NORG calculations, we study the ground state properties of a spin-$\frac{1}{2}$ impurity interacting with two helical liquids, based on the KM model defined in a finite ZGNR. We investigate how the Kondo couplings with both helical liquids influence the static and dynamic properties of the system. More specifically, we present a detailed study focusing on the NOs, the impurity spin polarization and susceptibility, as well as the Kondo screening cloud.

The occupancy number of the NOs is first analyzed. Our numerical results show that two ANOs with half occupancy emerge, whose number is equal to the number of the helical liquids interacting with the impurity. Structures of the two ANOs are further analyzed by their projections into both top and bottom edges in the ZGNR, which show difference around the coupling symmetry point.

It is shown that at the coupling symmetry point the impurity spin is maximally polarized and the susceptibility reaches the maximum. Furthermore, both the impurity spin polarization and the susceptibility are suppressed when the symmetry of the Kondo couplings with the two helical liquids is broken, demonstrating that the impurity tends to be screened without polarization when the Kondo couplings deviate well from the symmetry point. On the other hand, the impurity spin polarization decreases with the length of the ZGNRs increasing, meaning that the impurity spin polarization tends to vanish in the thermodynamic limit in the whole regimes of the controlling parameter.

It is illustrated that the impurity is correlated dominantly with the helical liquid with the larger Kondo coupling, namely, the Kondo cloud is mainly formed by the electrons in the helical liquid with the larger Kondo coupling, while it is equally formed by the electrons in the two helical liquids at the coupling symmetry point. On the other hand, the SOC effect breaks the symmetry in spatial distribution of the spin correlation, leading to the anisotropic Kondo cloud, in contrast to the isotropic Kondo cloud formed in a normal metal.

\begin{acknowledgments}

This work was supported by the National Natural Science Foundation of China (Grants No. 12104247 and No. 11934020). Computational resources were provided by the Physical Laboratory of High Performance Computing at Renmin University of China.

\end{acknowledgments}

\bibliography{SIKMTHL}

\end{document}